\documentstyle[11pt,aaspp4]{article}

\def\th{\hbox{\boldmath $\theta$}}

\def\Mpc{{\rm Mpc}}

\def\k{{\bf k}}

\def\n{\hat {\bf n}}
\def\hk{\hat {\bf k}}

\def\th{\hbox{\boldmath $\theta$}}

\begin{document}

\title{The Imprint of Lithium Recombination on the \\ Microwave Background
Anisotropies}

\author{Matias Zaldarriaga\altaffilmark{1} and Abraham
Loeb\altaffilmark{2}}

\altaffiltext{1}{Physics Department, New York University, 4
Washington Place, New York, NY 10003; matiasz@physics.nyu.edu}

\altaffiltext{2}{Astronomy Department, Harvard University, 60 Garden
Street, Cambridge, MA 02138; aloeb@cfa.harvard.edu}

\begin{abstract} 

Following Loeb (2001), we explore the imprint of the resonant 6708\AA~ line
opacity of neutral lithium on the temperature and polarization anisotropies
of the cosmic microwave background (CMB) at observed wavelengths of
$250$--$350\mu$m (0.9--1.2 THz).  We show that if lithium recombines in the
redshift range of $z=400$--$500$ as expected, then the standard CMB
anisotropies would be significantly modified in this wavelength band. The
modified polarization signal could be comparable to the expected
polarization anisotropies of the far--infrared background on sub--degree
angular scales ($\ell \ga 100$). Detection of the predicted signal can be
used to infer the primordial abundance of lithium, and to probe structure
in the Universe at $z\sim 500$.

\end{abstract}


\section{Introduction}

The latest measurements of the anisotropies in the cosmic microwave
background (CMB; see Halverson et al. 2001, Lee et al. 2001, Netterfield et
al. 2001) imply that the days of ``precision cosmology'' have already
arrived\footnote{A compilation of all experiments up to date can be found
at www.hep.upenn.edu/$^{\sim}$max/cmb/experiments.html}. Future ground and
balloon based experiments, in combination with the satellite missions
MAP\footnote{http://map.gsfc.nasa.gov} in 2001 and
Planck\footnote{http://astro.estec.esa.nl/Planck} in 2007, will test
current theoretical models to a sub--percent precision at photon
wavelengths $\ga 500\mu$m.

However, at far--infrared wavelengths of $\la 350\mu$m, Loeb (2001) has
recently shown that neutral lithium can strongly modify the CMB anisotropy
maps, through absorption and re-emission at its resonant 6708\AA~transition
from the ground state.  Lithium is expected to recombine in the redshift
interval $z\sim 400$--$500$ (Palla et al. 1995; Stancil et al. 1996, 1998).
Despite the exceedingly low lithium abundance\footnote{Note that by the
redshift of interest all the $^7$Be produced during Big--Bang
nucleosynthesis transforms to $^7$Li through electron capture, since $^7$Be
starts to recombine well before $^7$Li, owing to its significantly higher
ionization potential.} produced in the Big Bang, the resonant optical depth
(Sobolev 1946) after lithium recombination is substantial,
\begin{equation}
\tau_{_{\rm LiI}}[\lambda(z)]=1.00 f_{_{\rm LiI}}(z) \left({X_{_{\rm Li}}
\over 3.8\times 10^{-10}}\right) \left({1+z\over 500}\right)^{3/2} .
\label{eq:tau_Li}
\end{equation}
for an observed wavelength of $\lambda(z)=[6708$\AA$\times
(1+z)]=335.4\mu{\rm m}\times [(1+z)/500]$. Here, $X_{_{\rm Li}}\approx
3.8\times 10^{-10}$ is the latest estimate of the lithium to hydrogen
number density ratio (Burles et al. 2001), and $f_{_{\rm LiI}}(z)$ is the
neutral fraction of lithium as a function of redshift (Palla et al. 1995;
Stancil et al. 1996, 1998).  Loeb (2001) argued that resonant scattering
would suppress the original anisotropies by a factor of $\exp({-\tau_{_{\rm
LiI}}})$, but will generate new anisotropies in the CMB temperature and
polarization on sub--degree scales ($\ell \ga 100$), primarily through the
Doppler effect.  Observations at different far--infrared wavelengths could
then probe different thin slices of the early universe.

In this paper, we calculate in detail the effect of neutral lithium on both
the polarization and temperature anisotropies of the CMB.  \S 2 describes
the modifications we have made to the standard code, CMBFAST (Seljak \&
Zaldarriaga 1996), in order to calculate these anisotropies.  In \S 3, we
describe our results and compare them to the foreground noise introduced by
the far--infrared emission from galaxies and quasars. Finally, \S4
summarizes the main conclusions of this work.

Throughout the paper we adopt the LCDM cosmological parameters of
$\Omega_0\approx 0.25$, $\Omega_\Lambda=0.7$, $\Omega_b = 0.05$, and
$H_0\approx 70~{\rm km~s^{-1}~Mpc^{-1}}$, and units of $c=1$.

\section{Method of Calculation}

In order to compute the temperature and polarization fluctuations induced
by lithium scattering, we complement the standard Thomson opacity ($\dot
\tau_{_T}$) in CMBFAST by a new component ($\dot \tau_{_{\rm LiI}}$) 
which is assumed to have a narrow Gaussian shape in conformal time,
\begin{eqnarray}
\dot \tau &=&
\dot \tau_{_{\rm T}}(\eta) + \dot \tau_{_{\rm\ LiI}}(\eta,\nu)
\nonumber \\
&=& a n_e \sigma_T + \tau_{_{\rm LiI}} {
e^{-(\eta-\eta_{_{\rm LiI}})^2/2\sigma_\eta^2}
\over \sqrt{2\pi \sigma_\eta^2}} ,
\label{eq:tdot}
\end{eqnarray}
where, $\tau$ denotes optical depth, an overdot denotes a conformal time
derivative, $a$ is the expansion factor, $\eta$ is conformal time ($a\
d\eta=dt$), $n_e$ is the number density of free electrons, and $\sigma_T$
is the Thomson cross--section.  The lithium opacity depends both on time
and observed photon frequency, $\nu$. We characterize this opacity by three
parameters: the total optical depth $\tau_{_{\rm LiI}}$ given by equation
(\ref{eq:tau_Li}), and the central time ($\eta_{_{\rm LiI}}$) and width
($\sigma_\eta$) of the Gaussian peak. The peak is set to a redshift
$(1+z)=(\lambda/6708$ \AA$)$~, where $\lambda=c/\nu$ is the observed
wavelength.  The actual (non-instrumental) width of the Gaussian is
expected to be $[\Delta z/(1+z)]\sim 3\times 10^{-5}$ (Loeb 2001). Here we
adopt $\sigma_\eta/\eta\sim 10^{-2}$, as dictated by a fiducial detector's
band--width.

The lithium opacity introduces a frequency dependence to the CMB
anisotropies.  Next we show that different frequencies couple only through
the net drag force they provide on the baryons.  Our discussion follows the
notation of Ma \& Bertschinger (1995).

We consider a density perturbation of comoving wave-vector $\k$.  The
photon distribution function can then be expanded to first order in the
perturbation amplitude as,
\begin{eqnarray}
f(\k,\n,\nu,\eta)&=&f_0(\nu)(1+\Psi(\k,\n,\nu,\eta)) , \nonumber \\
\Psi(\k,\n,\nu,\eta) &=& \sum_{l=0}^{\infty} (-i)^l (2l+1)
\Psi_l(k,\nu,\eta) P_l(\hk\cdot\n).  
\label{expan}
\end{eqnarray}
where $f_0(\nu)=2/h^3(e^{h\nu/kT_0}-1)$, and where $f$ is a function of
perturbation wave-vector ($\k$), time ($\eta$), and frequency ($\nu$) and
propagation direction ($\n$) of the photon. In the second line of equation
(\ref{expan}), we have expanded the angular dependence of the distribution
function in Legendre polynomials.

The distribution function $f(\k,\n,\nu,\eta)$ satisfies the Boltzmann
equation.  Since the Thomson cross--section is independent of frequency,
the standard approach integrates the Boltzmann equation over photon
frequencies and uses only one Boltzmann hierarchy to evolve the photon
distribution function. In our case, the frequency dependence of the lithium
cross--section implies that one should solve the hierarchy for different
frequencies.  Since the frequencies are coupled, one needs to follow all
frequencies simultaneously, as done by Yu et al. (2001) for Rayleigh
scattering. The coupling between different frequencies does not originate
directly from the scattering term because lithium scattering does not
change the photon frequency. As we show next, the coupling arises from the
drag force on the baryons. Binary particle collisions allow the baryons to
behave as a single fluid (Loeb 2001) which is subject to the sum of the
forces applied by photons at all frequencies.  In the limit of no drag
force, the different frequencies decouple.

Next, we define the relative density contrast ($\delta_\gamma$) and
velocity divergence of the photon fluid ($\theta_\gamma$) as,
\begin{eqnarray}
\delta_{\gamma}&=& (a^4 \bar \rho_{_{\rm CMB}})^{-1} 
{\int d^3 \nu \  \nu f_0(\nu) \
\Psi_0(k,\nu,\eta)}, 
\nonumber \\
\delta\theta_{\gamma}(\nu)&=& (a^4 \bar \rho_{_{\rm CMB}})^{-1} \ \ 
{3\over 4} k { \Psi_1(k,\nu,\eta)},  \nonumber \\
\theta_{\gamma}&=& \int d^3 \nu \  \nu f_0(\nu) \
\delta\theta_\gamma(\nu) .
\end{eqnarray}
With these definitions we can write the equation for the velocity
divergence of the baryon fluid ($\theta_{b}=k v_b$),
\begin{eqnarray}
\dot \theta_{b}&=& -{\dot a \over a} + c_s^2 k^2 \delta_b 
+ {4 \bar \rho_{_{\rm CMB}} \over 3 \bar \rho_b} 
\left[ \dot \tau_{\rm T}
(\theta_\gamma - \theta_b) +  
\int d^3 \nu \  {\dot \tau}_{_{\rm\ LiI}}(\eta,\nu) \ \nu f_0(\nu)  
(\delta\theta_\gamma-\delta\theta_b)\right] ~,\nonumber \\
\delta\theta_b&\equiv & (a^4 \bar \rho_{_{\rm CMB}})^{-1}\ { \theta_b},
\label{baryons}
\end{eqnarray}
where $c_s$ is the sound speed of the baryons, $\bar \rho_{_{\rm CMB}}$ and
$\bar \rho_b$ are the mean energy densities of the CMB and the baryons, and
$\delta_b$ is the baryon overdensity.  The last term in the square brackets
of equation (\ref{baryons}) is responsible for the coupling between
different photon frequencies.  However, Figure 1 of Loeb (2001) implies
that we may ignore the drag force on the baryons (due to either Thomson or
lithium scattering) at the redshifts where the lithium opacity becomes
important ($z\la 500$).  Consequently, we may solve the Boltzmann equation
separately for each photon frequency by explicitly neglecting the transfer
of momentum from the photons to the baryons due to lithium scattering.
Even though momentum conservation is not strictly satisfied in this
approach, the remaining correction is expected to be negligible.

The last change that we introduce to CMBFAST involves polarization. The
cross--section for lithium scattering has a different dependence from
Thomson scattering on both scattering angle and Stokes parameters. The
scattering matrix for $(I_\parallel,I_\perp,U)$ (where the parallel and
perpendicular directions are defined relative to the scattering plane) can
be decomposed into two parts,
\begin{equation}
{3\over 2} E_1 \left( \begin{array}{ccc}
\cos^2 \Theta & 0 & 0 \\
0 & 1 & 0 \\
0 & 0 & \cos \Theta \\
\end{array} \right) + {1\over 2} E_2
\left( \begin{array}{ccc}
1 & 1 & 0\\
1 & 1 & 0\\
0 & 0 & 0
\end{array} \right), 
\end{equation}
where $\Theta$ is the scattering angle. The first term is the usual Thomson
(or Rayleigh) scattering matrix multiplied by a factor $E_1$, while the
second term, which is proportional to $E_2$, does not generate polarization
and is isotropic in angle.  The two amplitudes depend on the quantum
numbers of the resonant states, and satisfy $E_1+E_2=1$ (Chandrasekar
1960). For the transition between the ground state ($2S$) and first excited
state ($2P$) of lithium, we get $E_1=1/3$ (Hamilton 1947; Chandrasekar
1960). Interference between the $2^2S$--$2^2P^0_{1/2}$ transition and the
$2^2S$--$2^2P^0_{3/2}$ transition (Stenflo 1980) has a negligible effect on
the polarization.  This follows from the fact that $\tau_{_{\rm LiI}}\sim
1$; and for a given lithium atom, a photon will likely scatter when its
frequency is separated from the line center by less than the natural width
(37 MHz/$4\pi$), which is much smaller than the frequency separation
between these transitions (10 GHz).

\section{Results}

\begin{figure}[tb]
\centerline{\epsfxsize=9cm\epsffile{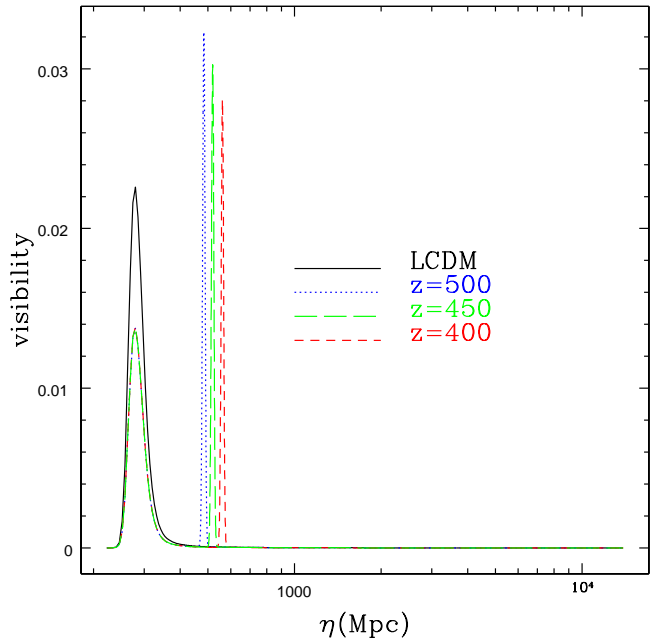}}
\caption{Visibility function for models with $\tau_{_{\rm LiI}}=0.5$ and
LCDM. We show three examples, corresponding to resonant scattering at
different redshifts, $z=400$, $z=450$ and $z=500$. The present--day value
of the conformal time is $\eta_0=1.39\times 10^4$ Mpc.}
\label{fig8}
\end{figure}

The significance of the new opacity component can be assessed from the
visibility function $\Upsilon(\eta)$.  This function provides the
probability distribution for the time of last scattering of the photons
observed today at a conformal time $\eta_0$,
\begin{eqnarray}
\Upsilon(\eta) &=&\dot \tau(\eta) e^{-\tau(\eta)}, \nonumber \\ \tau(\eta) &=&
\int_\eta^{\eta_0} d\eta^{\prime} \dot \tau(\eta^{\prime}).
\end{eqnarray}
In Figure \ref{fig8} we show the visibility functions for some of the
models we consider later. 

The observed anisotropies have two separate contributions, one from the
standard last scattering surface at hydrogen recombination (decoupling),
which is suppressed by $e^{-\tau_{_{\rm LiI}}}$, and a second new
contribution that is generated by lithium scattering at lower redshifts. In
the following sub-sections we will characterize this new contribution to
the temperature and polarization anisotropies.

\subsection{Temperature and Polarization Anisotropies} 

\begin{figure}[tb]
\centerline{\epsfxsize=9cm\epsffile{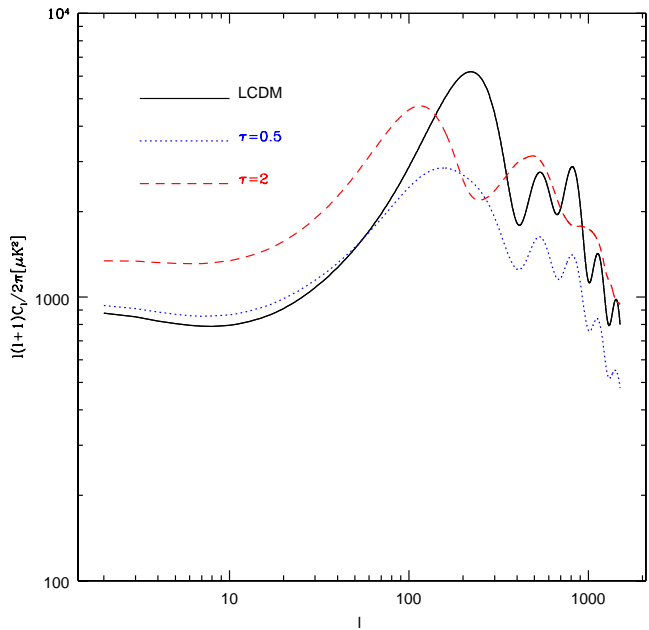}}
\caption{Temperature power spectra for the standard calculation (LCDM
model), and two other models with lithium optical depths of $\tau_{_{\rm
LiI}}=0.5$ or $\tau_{_{\rm LiI}}=2.0$ at $z=500$, for observations in a
narrow--band around 335$\mu$m.  The width of the $\dot \tau_{_{\rm LiI}}$
Gaussian in Eq. (\ref{eq:tdot}) is $\sigma_\eta/\eta=0.01$.}
\label{fig1}
\end{figure}

Figures \ref{fig1} and \ref{fig2} show the predicted power spectra for the
temperature and polarization anisotropies of the CMB at an observed
wavelength of $\lambda=335.4 \mu$m, corresponding to lithium scattering at
$z=500$.  The Stokes parameters are measured in $\mu K$. The figures
compare the spectrum of fluctuations in standard LCDM (no lithium
scattering) with two other models, each having a peak in $\dot \tau_{_{\rm
LiI}}$ at a redshift of $z=500$, but with a total optical depth of either
$\tau_{_{\rm LiI}}=0.5$ or $\tau_{_{\rm LiI}}=2.0$.  Since at long
wavelengths, $\lambda \ga 500 \mu$m, the LCDM fluctuations are not altered,
precise mapping of these fluctuations by the MAP or Planck satellites will
provide a reference power spectrum against which the lithium distortion can
be measured.

Figure \ref{fig1} shows that for $\tau_{_{\rm LiI}}=0.5$ the small scale
fluctuations are dominated by the suppressed anisotropies from
recombination, resulting in a power spectrum which is similar in shape to
that of the primary anisotropies but suppressed in amplitude.  However, the
$\tau_{_{\rm LiI}}=2$ case is very different. Here, the anisotropies are
actually larger for many $l$'s than those expected without the lithium
scattering, having a different functional dependence on $l$ than the
standard case. The $e^{-\tau_{_{\rm LiI}}}$ suppression of the original
anisotropies is sufficient to make them sub-dominant relative to the newly
generated anisotropies at $z=500$.

\begin{figure}[tb]
\centerline{\epsfxsize=9cm\epsffile{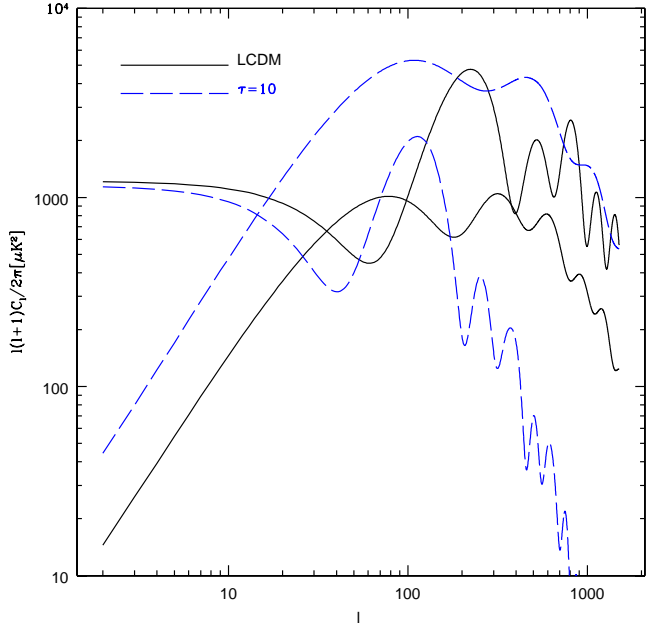}}
\caption{Velocity and temperature contribution to anisotropies for LCDM and
for a model with $\tau_{_{\rm LiI}}=10$ at $z=500$. The curves that
approach a finite value at low $l$ are the temperature contributions, and
the curves that approach zero at low $l$ describe the velocity
contributions. }
\label{fig9}
\end{figure}

\begin{figure}[tb]
\centerline{\epsfxsize=9cm\epsffile{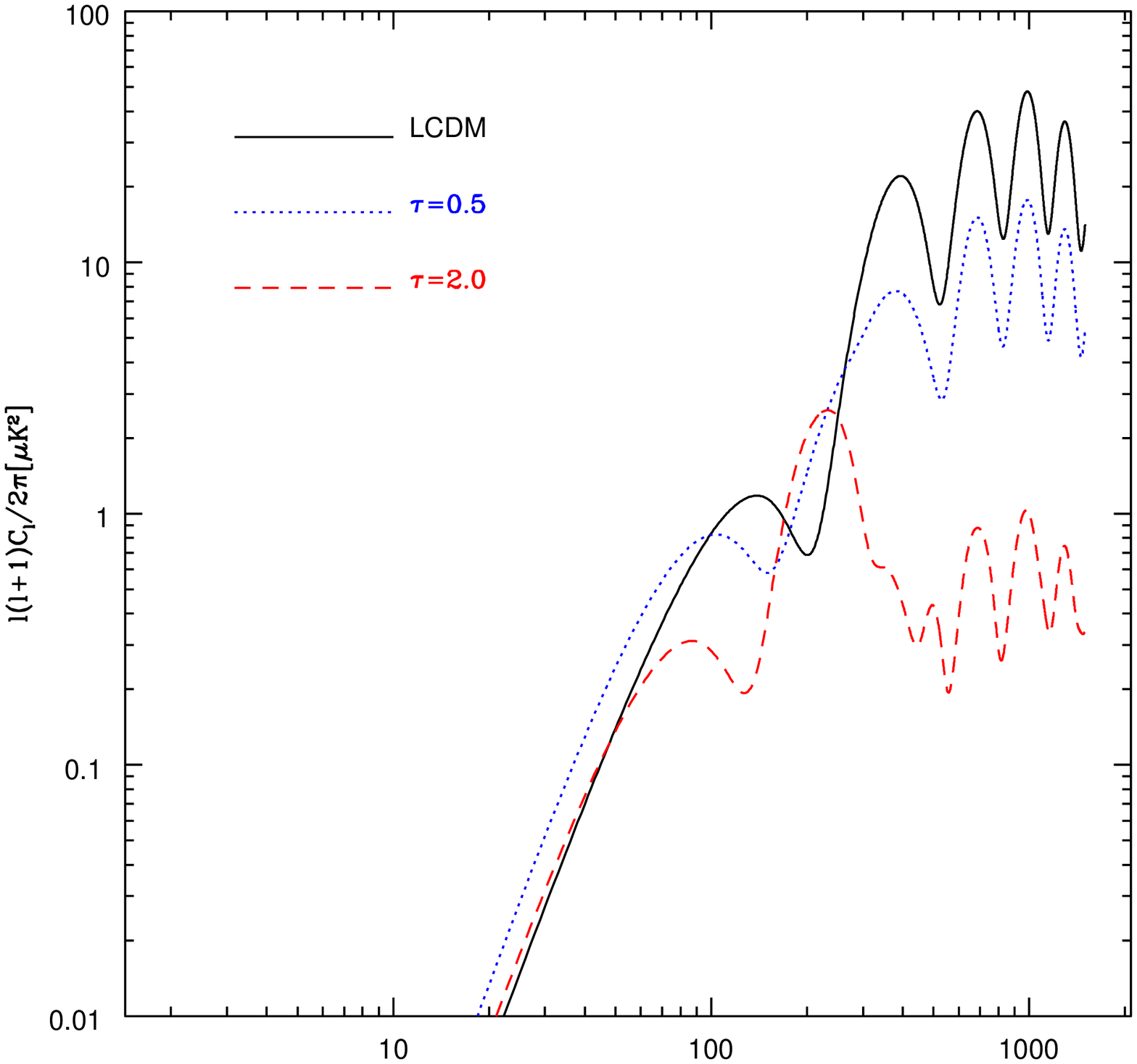}}
\caption{Polarization power spectra. Models are the same as
in Fig. \ref{fig1}.}
\label{fig2}
\end{figure}

There is an interesting difference between the primary anisotropies and
those created by lithium scattering. In order to explain it, we introduce
the integral solution for the temperature anisotropies,
\begin{equation}
\left({\Delta T \over T} + \psi\right)(k,\mu,\eta_0)=\int_0^{\eta_0} d\eta
\left[(\delta_{\gamma}/4 + \psi + \mu v_b) \Upsilon(\eta) + (\dot \psi + \dot
\phi) e^{-\tau(\eta)}\right] e^{ik\mu(\eta-\eta_0)},
\end{equation}
where $\mu$ is the cosine of the angle between the wave-vector and the
direction of observation and $\phi$ and $\psi$ are the two gravitational
potentials defined by the perturbed metric, $ds^2 = a^2(\eta)
[-(1+2\psi)d\eta^2 + (1-2\phi) dx_i dx^i]$.  While the contribution from
recombination is dominated by the monopole term, ($\delta_\gamma/4+\psi$),
the lithium anisotropies are dominated by the peculiar velocity term
($\propto v_b$) on most scales.  We illustrate this in Figure \ref{fig9}
where we show the monopole and velocity contributions to the anisotropies
in both the standard LCDM model and in a model that has a large optical
depth $\tau_{_{\rm LiI}}=10$ at $z=500$.  We chose such a large optical
depth in order to suppress the original contribution from decoupling.  The
figure clearly shows that the anisotropies are dominated by the monopole
term for the standard LCDM model for almost all values of $l$, while the
opposite is true for the $\tau_{_{\rm LiI}}=10$ model.  The new
anisotropies are dominated by the monopole term only at very low
multipoles, $l\la 20$.

We can easily explain why the monopole no longer dominates for the lithium
contribution.  After recombination the monopole term decays by the free
streaming of the photons while the velocity of the baryons continues to
grow as they fall into the dark matter potential wells. For a perturbation
mode of wave-vector $k$, the monopole term at conformal time $\eta$ after
recombination, $\eta>\eta_{\rm rec}$, is approximately given by,
\begin{equation}
(\delta_\gamma/4 + \psi)(\eta)=(\delta_\gamma/4 + \psi)(\eta_{\rm rec})
j_0\left[ k(\eta-\eta_{\rm rec})\right] ,
\label{monopole}
\end{equation}
where $j_0(x)$ is the spherical Bessel function.  Equation (\ref{monopole})
shows that the monopole term is small for $k(\eta-\eta_{\rm rec}) \gg 1$
because of the decay in the Bessel function when its argument is
large. Figure \ref{fig8} shows that the new peak of the visibility function
occurs at $\eta\sim 500\ \Mpc$ while $\eta_{\rm rec}\sim 300\ \Mpc$.  We
can translate the spatial wavenumber $k$ to angular scale using the
conformal distance to the new peak in the visibility function, $d =
(\eta_0-\eta) \approx \eta_0$. We find that $k(\eta-\eta_{\rm rec})\sim
1.6\times 10^{-2} l$, which explains why the monopole term is suppressed
for $l \gg 60$.

While the monopole term decays between recombination and the lithium
scattering surface, the velocity grows and thus produces anisotropies that
are larger than those generated at decoupling in the $\tau_{_{\rm LiI}}=2$
case.

The physics of the polarization anisotropies is different from that of the
temperature anisotropies. Since polarization is generated by the quadrupole
moment, there are two competing effects that need to be considered.  On the
one hand, the quadrupole anisotropies are small at recombination, since
they are suppressed relative to the velocity fluctuations by a factor $k\
\delta\eta$, where $\delta\eta$ is the width of the last scattering surface
at recombination (Zaldarriaga \& Harari 1995). In the new scenario the
quadrupole is able to grow during the free streaming period between
recombination and $z\sim 500$. This naturally leads to an increase in the
polarization signal. The same effect increases the polarization
anisotropies on large scales in models with a substantial optical depth to
Thomson scattering after the universe reionizes (Zaldarriaga 1997).  On the
other hand, due to the nature of resonant--line scattering (Hamilton 1947;
Chandrasekhar 1960), only $1/3$ of the cross--section generates
polarization out of this quadrupole, while $2/3$ produces unpolarized
radiation ($E_1=1/3,\ E_2=2/3$).  Although the quadrupole is bigger at
$z\sim 500$ than at recombination, the newly generated polarization is not
as large. For example, Figure \ref{fig2} shows that even for $\tau_{_{\rm
LiI}}=2$, the polarization at high multipoles, $l\sim 1000$, is dominated
by the suppressed signal from decoupling.


\subsection{Cross--correlations and Difference Maps}

Let us next consider two temperature anisotropy maps, obtained from
observations at different wavelengths, $T^a(\th)$ and $T^b(\th)$. We define
the following spectra in multipole space,
\begin{eqnarray}	
\langle T_l^a T_{l^\prime}^a\rangle &=&\delta_{ll^{\prime}}C_l^{aa} ,
\nonumber \\ \langle T_l^b T_{l^\prime}^b\rangle
&=&\delta_{ll^{\prime}}C_l^{bb} ,\nonumber \\ \langle T_l^a
T_{l^\prime}^b\rangle &=&\delta_{ll^{\prime}}C_l^{ab},
\end{eqnarray}	
with $\delta_{ll^{\prime}}$ being the Kronecker delta.  Similar expressions
can be written for $E$ and $B$--type polarization.  We then define the
cross--correlation coefficient,
\begin{equation}
CC_l={C_l^{ab}\over \sqrt{C_l^{aa}C_l^{bb}}}.
\end{equation}
If $CC_l\rightarrow 1$ across a range of $l$'s, then the two maps are
scaled versions of each other over that range.

\begin{figure}[tb]
\centerline{\epsfxsize=9cm\epsffile{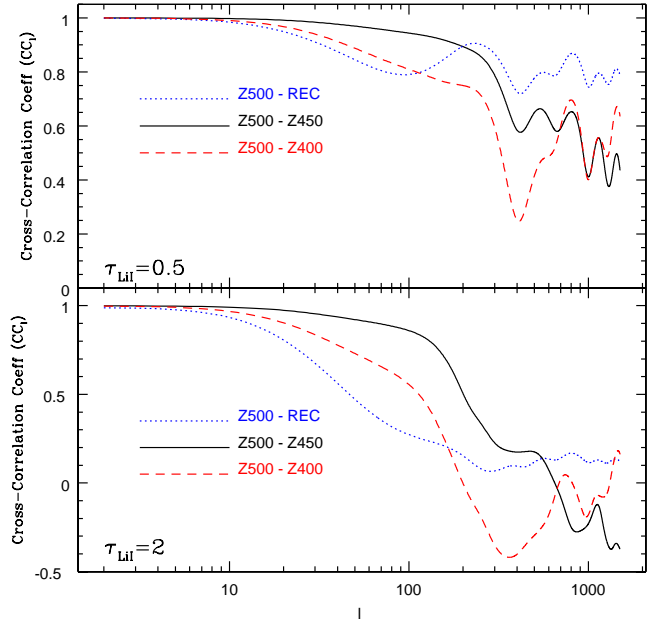}}
\caption{
Correlation coefficients for temperature anisotropies at three
different wavelengths corresponding to lithium resonances at $z=400$
(268$\mu$m), $z=450$ (302$\mu$m) and $z=500$ (335$\mu$m). The dotted lines
show the correlation coefficient between the standard (long wavelength)
anisotropies and those with lithium scattering at $z=500$; the solid line
shows the correlation between the anisotropies for 302 $\mu$m and 335$\mu$m
($z=500$ and $z=450$); and the dashed lines correspond to 335$\mu$m and
268$\mu$m ($z=500$ and $z=400$).  The upper panel shows the case of
$\tau_{_{\rm LiI}}=0.5$ and the lower panel shows the case of $\tau_{_{\rm
LiI}}=2.0$.}
\label{fig4}
\end{figure}

\begin{figure}[tb]
\centerline{\epsfxsize=9cm\epsffile{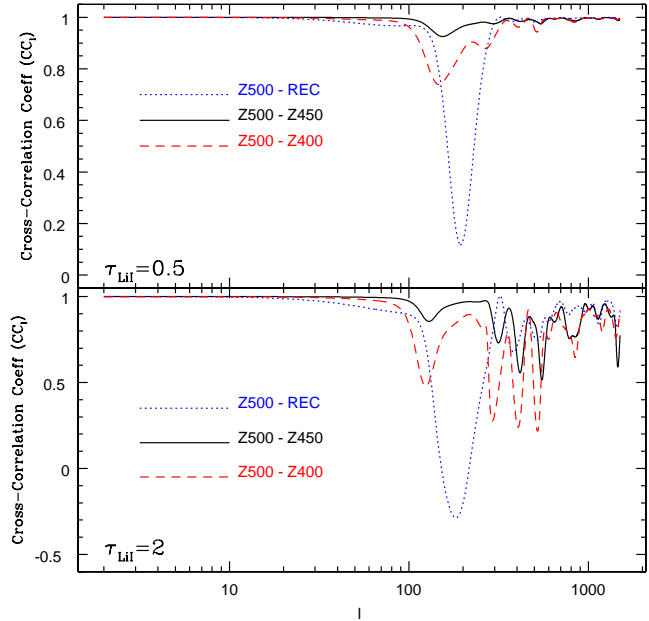}}
\caption{Correlation coefficient for polarization. The upper panel shows
the case of $\tau_{_{\rm LiI}}=0.5$, and the lower panel shows the case of
$\tau_{_{\rm LiI}}=2.0$}
\label{fig5}
\end{figure}

Figures \ref{fig4} and \ref{fig5} show the correlation coefficients between
maps of the anisotropies at different observed wavelengths, labelled by the
redshift of the peak in the visibility function. They correspond to the
correlation coefficient in three cases: (i) between the primary
anisotropies and those measured at $\lambda=335.4\ \mu$m ($z=500$); (ii)
between measurements at $\lambda=335.4\ \mu$m and $\lambda=301.8\ \mu$m
($z=450$); and (iii) between $\lambda=335.4\ \mu$m and $\lambda=268.3\
\mu$m ($z=400$).

The temperature correlation coefficient between the original map of the
primary anisotropies and a map at 335.4 $\mu$m can be understood as
follows.  For low $l$, $CC_l \approx 1$ because the anisotropies are
produced by long wavelength (small $k$) modes and the difference in
conformal time between recombination and $z\sim 500$ (which we denote by
$d_{\rm rec-500}$) is smaller than the perturbation wavelength , i.e. $(k\
d_{\rm rec-500} << 1)$. The fact that there are two different scattering
surfaces does not make a difference for these modes.

On small scales, the situation is more complicated. The component of the
anisotropies from decoupling, which is suppressed by a factor
$e^{-\tau_{_{\rm LiI}}}$, drives the cross--correlation coefficient to
unity, because its suppression results merely in rescaling.  However, the
newly generated anisotropies are uncorrelated with those coming from
decoupling if they are produced by wavelengths that are smaller than
$d_{\rm rec-500}$, i.e. if $k\ d_{\rm rec-500} \sim (d_{\rm rec-500}/d_{\rm
rec}) l \sim 1.6\times 10^{-2} l \gg 1$. The newly generated anisotropies
tend to drive the cross--correlation coefficient to zero on small scales.

In Figure \ref{fig4} we also show the cross--correlations between maps at
observed wavelengths for which $\dot \tau_{_{\rm LiI}}$ peaks at $z=400$,
$z=450$, and $z=500$. For a given angular scale, the anisotropies generated
by the second peak of the visibility function are different only if the
conformal distance between the two peaks is much larger than the wavelength
of the perturbation producing the anisotropies on that scale. This explains
why the departure from $CC_l=1$ occurs on smaller scales for the
correlation between $z=500$ and $z=450$ than for that between $z=500$ and
$z=400$. Also, if the anisotropies in the two maps have a substantial
contribution from decoupling, then the correlation coefficient will not
approach zero even on small scales where the newly generated anisotropies
in the two maps are uncorrelated. The significance of this primary
contribution from decoupling is suppressed when $\tau_{_{\rm LiI}}$
increases, as illustrated by the $\tau_{_{\rm LiI}}=0.5$ and $\tau_{_{\rm
LiI}}=2.0$ panels of Figure \ref{fig4}.

The correlation coefficient for polarization which is shown in Figure
\ref{fig5}, can be explained along similar lines. On small scales, the
polarization anisotropies are dominated by the suppressed contribution from
recombination, and so $CC_l \rightarrow 1$.  The dominance of the new
anisotropies over the primary ones at $l\sim 200$, makes the correlation
coefficient between primary and $z=500$ deviate away from unity around that
scale. However, among neighboring frequencies the cross--correlation
approaches unity near $l\sim 200$, because the distance between the peaks
of the visibility function is not sufficient to decorrelate the
contributions from the relevant $k$ modes.

\begin{figure}[tb]
\centerline{\epsfxsize=9cm\epsffile{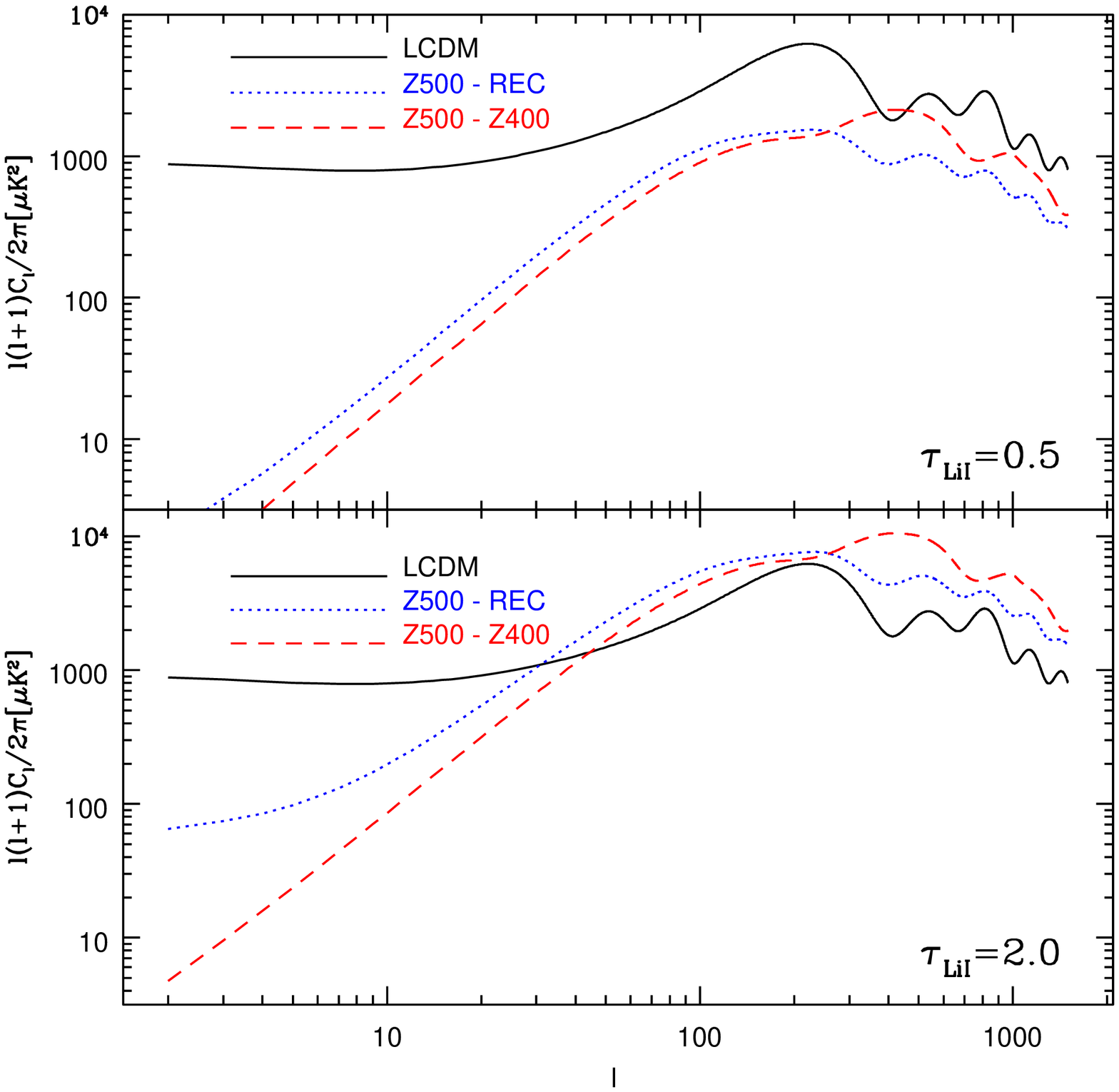}}
\caption{Difference power spectra for temperature anisotropies. In the
upper panel $\tau_{_{\rm LiI}}=0.5$, and in the lower panel 
$\tau_{_{\rm LiI}}=2.0$. }
\label{fig6}
\end{figure}

\begin{figure}[tb]
\centerline{\epsfxsize=9cm\epsffile{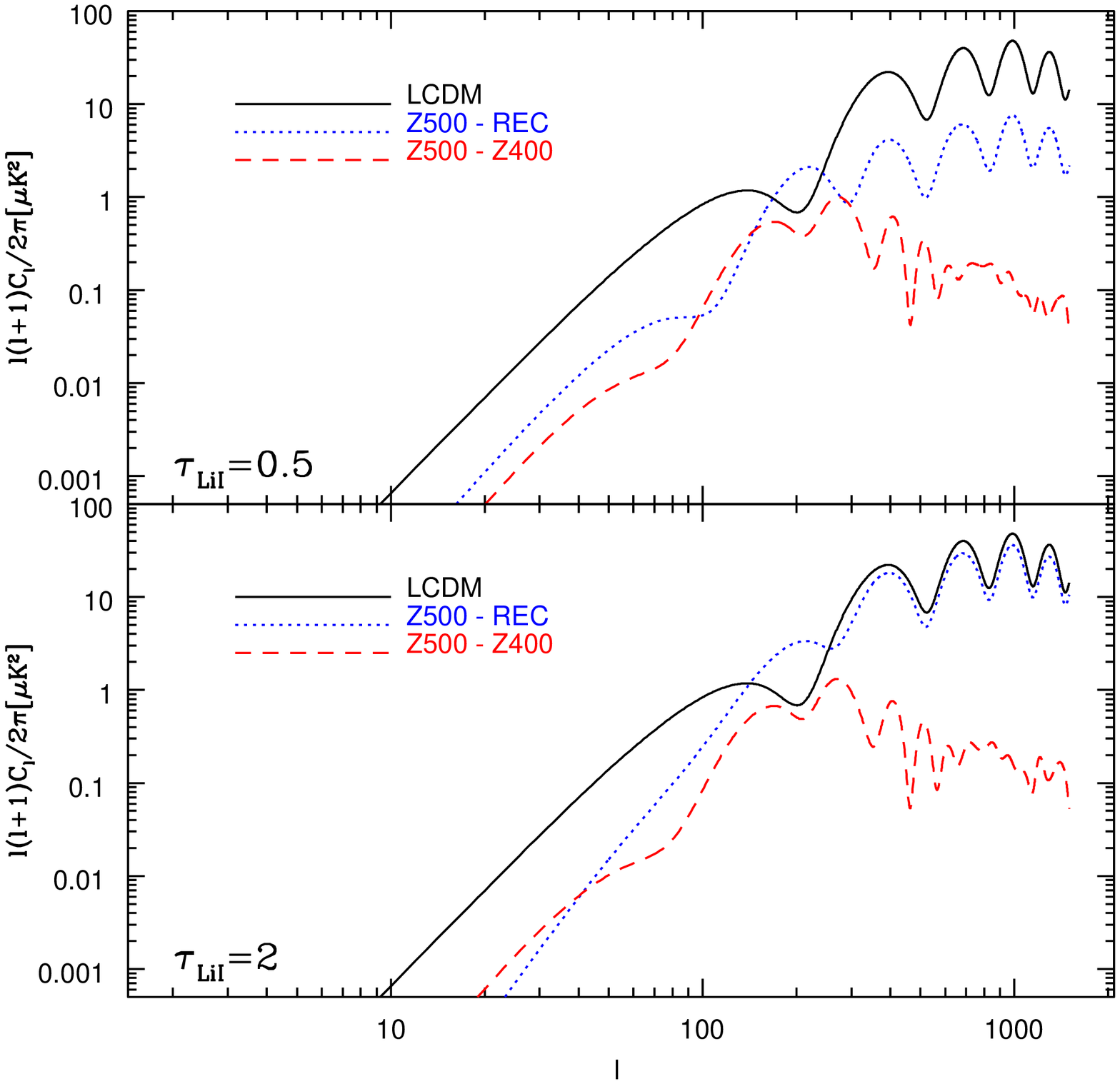}}
\caption{Difference spectra for polarization. In the upper panel
$\tau_{_{\rm LiI}}=0.5$ and in the lower panel $\tau_{_{\rm LiI}}=2.0$.}
\label{fig7}
\end{figure}

In Figures \ref{fig6} and \ref{fig7} we show the difference power spectra
for two maps at different observed wavelengths. We define $\delta
T(\th)=T^a(\th)-T^b(\th)$ and
\begin{equation}	
\langle\delta T_l\delta T_{l^\prime}\rangle
=\delta_{ll^{\prime}}C_l^{\rm diff}. 
\end{equation}	
A similar expression can be written for the polarization.  The difference
spectrum provides the new signature due to lithium, since the MAP or Planck
satellites will measure with high precision the anisotropies at long photon
wavelengths.

The power spectra of the difference maps, shown in Figure \ref{fig6} for
the temperature and Figure \ref{fig7} for the $E$--type polarization, are
consistent with our interpretation of the cross--correlation spectra
between maps. On large scales, the difference between the temperature maps
becomes very small. The power in the $Z500$--$Z400$ difference map peaks at
$l\sim 450$, while the power for the {\it recombination}--$Z500$ difference
peaks at $l\sim 250$. The amplitude of the difference increases as the
optical depth increases. When $\tau_{_{\rm Li I}}$ is large, the difference
spectrum for the {\it recombination}--$Z500$ polarization almost coincides
with that of the primary anisotropies, as a result of the fact that the
primary anisotropies dominate over the newly generated ones.

\subsection{Comparison to the Far--Infrared Foreground}

The main difficulty in measuring the lithium imprint on the CMB
anisotropies is the contamination by the far--infrared background
(FIB). There has been no detection to date of the anisotropies in this
background, and so we have to rely on theoretical estimates (Haiman \& Knox
2000; Knox et al. 2001).

In order to properly combine the contributions from the CMB and FIB
anisotropies we express intensities in terms of the equivalent
Rayleigh--Jeans temperature, $T_{\rm RJ}$ (in $\mu$K).  We start by
comparing the temperature anisotropies. The total fluctuation amplitude is
given by,
\begin{equation}
\Delta T_{\rm RJ}=\Delta T_{\rm RJ}^{\rm CMB}+\Delta T_{\rm RJ}^{\rm FIB}
=\Delta T_{\rm RJ}=T_{\rm RJ}^{\rm CMB} \left({\Delta T_{\rm RJ}\over
T_{\rm RJ}}\right)_{\rm CMB} + T_{\rm RJ}^{\rm FIB} \left({\Delta T_{\rm
RJ}\over T_{\rm RJ}}\right)_{\rm FIB} .
\end{equation}

The ratio between the CMB intensity and the central value for the inferred
intensity of the FIB (Fixsen et al. 1998; see the central dot--dashed curve
in Figure 2 of Haiman \& Knox 2000) is of order unity for a lithium
scattering redshift $z\sim 500$,
\begin{equation}
{T_{\rm RJ}^{\rm CMB}\over T_{\rm RJ}^{\rm FIB}} \approx \left({500\over
1+z}\right) \exp \left\{15.78 \left(1- {500\over 1+z}\right)\right\} .
\end{equation}
As noted by Loeb (2001), the temperature fluctuations in the Wien tail
translate to intensity fluctuations, $Delta I_\nu$ (in ${\rm
erg~s^{-1}cm^{-2}sr^{-1}Hz^{-1}}$), of a much larger contrast,
\begin{equation}
\left({\Delta T_{\rm RJ}\over T_{\rm RJ}}\right)_{\rm CMB}\equiv {\Delta
I_\nu\over I_\nu}= \left({d\ln I_\nu \over d\ln T}\right){\Delta T\over T}=
\left({h\nu\over kT}\right){\Delta T\over T}= 15.78 \times \left({500\over
1+z}\right){\Delta T\over T}~~,
\end{equation}
where we have substituted $I_\nu(T)\propto \exp(-h\nu/kT)$ and $T=2.725$ K
(Mather et al. 1999).  The anisotropy amplitude shown in Figure \ref{fig6}
depends on $\tau_{_{\rm LiI}}$, but roughly implies $ (\Delta T_{\rm
RJ}/T_{\rm RJ})_{\rm CMB}\sim 3\times 10^{-4}[500/(1+z)]$.  Haiman \& Knox
(2000) and Knox et al. (2001) estimate, $(\Delta T_{\rm RJ}/T_{\rm
RJ})_{\rm FIB}=0.05$--$0.1$.  The FIB anisotropies peak at an $l$ of a few
hundred but the peak is very broad. The anisotropies in the FIB are
relatively large since they originate from clustering of sources at low
redshifts, $z\sim 1$.

We conclude that if $50\%$ of the lithium ions recombine at $z\sim 500$
then
\begin{equation}
{\Delta T_{\rm RJ}^{\rm CMB}\over \Delta T_{RJ}^{\rm FIB}} \approx 3 \times
10^{-3} \left({500\over 1+z}\right)^2 \exp \left\{15.78 \left(1- {500\over
1+z}\right)\right\} .
\label{eq:t_cmb}
\end{equation}
Since the CMB contribution is sub-dominant, it is essential to exploit the
different frequency dependence of the FIB and CMB anisotropies in order to
subtract the FIB contribution with high precision. This might be possible
on small angular scales, where the temperature anisotropies generated by
lithium scattering at different redshifts are uncorrelated, as indicated by
Figures \ref{fig4} and \ref{fig6}.  Also, since the FIB is produced by
point sources, observations with high angular resolution can resolve the
sources and remove them individually.

Next we consider polarization. While the intensity anisotropies of the FIB
are dominated by the clustering of the sources rather than by Poisson
fluctuations, the situation is different for the polarization
anisotropies. We assume that each source is polarized with a degree of
polarization $\epsilon$, and that the polarizations of different sources
are uncorrelated. Under these assumptions, it is easy to show that both the
$E$ and $B$--polarization spectra share the same amplitude of
\begin{equation}
\epsilon^2/2 \times C_{Tl}^{\rm Poisson},
\end{equation}
where $C_{Tl}^{\rm Poisson}$ is the Poisson contribution to the temperature
anisotropies. The Poisson fluctuations can be calculated based on the SCUBA
source counts, and are shown as the light solid line in Figure 3 of Haiman
\& Knox (2000).

The Poisson component of the FIB anisotropies is approximately a factor of
$\sim 100$ smaller in power than the total temperature anisotropies at
$l\sim 100$--$200$.  If we adopt an average value of $\epsilon = 0.4\%$
(Jones 1993), then the $E$--polarization is lower by a factor
$(0.4\%/\sqrt{2}) \times 0.1 = 2.8 \times 10 ^{-4}$ than the intensity
fluctuations.

On the other hand, Figures \ref{fig1} and \ref{fig2} imply that the CMB is
approximately 10 \% polarized for $l\ga 200$. We therefore get that as long
as lithium recombines around $z\sim 500$,
\begin{equation}
{\Delta E_{\rm RJ}^{\rm CMB}\over \Delta E_{\rm RJ}^{\rm FIB}} \approx 1.1
\times \left({\epsilon \over 0.4\%}\right) \left({500\over 1+z}\right)^2
\exp \left\{15.78 \left(1- {500\over 1+z}\right)\right\} .
\label{eq:e_cmb}
\end{equation}

A property which could be very useful in experimental attempts to isolate
the signal is that the polarization signature due to lithium has only an
$E$--type component, while the FIB polarization has an equal amplitude in
both the $E$ and $B$--type components.  The $B$--type polarization could
therefore be used to monitor the FIB contamination and could play an
essential role in the subtraction of the FIB.

Depending on the nature of the sources responsible for the FIB and their
luminosity function it may eventually become possible to resolve most of
the FIB through high--resolution observations at different
wavelengths. This approach is used, for example, in observational studies
of the Sunyaev--Zeldovich effect, where much of the contribution from
discrete foreground sources is subtracted out through deep,
high--resolution observations at either radio or optical--infrared
wavelengths.  At the present time there are no available source counts in
the wavelength range we consider here. Closest in wavelength are source
counts from the SCUBA instrument (see, e.g. Fig. 2 in Borys et
al. 2000). If most of the FIB could be resolved, the task of detecting the
effect of lithium would become easier as the overall level of contamination
would be drastically reduced. Future studies of the FIB will determine
whether this reduction is feasible.

\section{Conclusions}

We have shown that if more than half of the lithium ions recombine by
$z\sim 500$, then the temperature and polarization anisotropies of the
CMB would be strongly altered at an observed wavelength of $335 \mu$m
(see Figs. \ref{fig1} and \ref{fig2}). For high multipoles $l\gg 10$,
the change is dominated by two contributions: (i) the Doppler
anisotropies induced at the sharp lithium scattering surface; and (ii)
the uniform $\exp\{-\tau_{_{\rm LiI}}\}$ suppression of the primary
anisotropies which were generated at hydrogen recombination
(decoupling).  Maps taken at wavelengths that are different by only
$\sim 10\%$ are expected to have significant differences (see
Figs. \ref{fig6} and \ref{fig7}).

The above signals are superimposed on top of the far infrared background
(FIB).  Our estimates imply that the lithium imprint on the CMB
polarization should be comparable to that provided by the FIB
(Eq. \ref{eq:e_cmb}). Detection is more difficult for the temperature
anisotropies (Eq. \ref{eq:t_cmb}).

The wavelength range we explored overlaps with the highest frequency
channel of the Planck mission ($352\mu$m) and with the proposed
balloon--borne Explorer of Diffuse Galactic
Emissions\footnote{http://topweb.gsfc.nasa.gov} (EDGE) which will survey
1\% of the sky in 10 wavelength bands between $230$--$2000\mu$m with a
resolution ranging from $6^\prime$ to $14^\prime$ (see Table 1 in Knox et
al. 2001). In order to optimize the detection of the lithium signature on
the CMB anisotropies, a new instrument design is required, with multiple
narrow bands ($\Delta \lambda/\lambda\la 0.1$) at various wavelengths in
the range $\lambda=250$--$350\mu$m. The experiment should cover a
sufficiently large area of the sky so as to determine reliably the
statistics of fluctuations on degree scales. In order to minimize
contamination from the FIB, the detector should be sensitive to
polarization. For reference, the experiment should also measure the
anisotropies at shorter wavelengths where the FIB dominates.  In order to
detect the effect of lithium, high signal-to-noise maps of the primordial
CMB should be made for the same region of the sky. Most likely, those maps
will become available from future CMB missions such as Planck. A strategy
for eliminating the contribution from the brightest FIB sources may also be
needed.

The resonant optical depth depends sensitively on the primordial lithium
abundance and the recombination history of lithium.  More detailed
calculations of lithium recombination will be done in a forthcoming paper
(Dalgarno, Loeb, \& Stancil 2001). Detection of the lithium signature will
also allow to calibrate the primordial lithium abundance, which is a
sensitive indicator of the mean value and the clumpiness in the baryon
abundance during Big Bang nucleosythesis. The lithium abundance in nearby
stars is subject to large astrophysical uncertainties (Burles et al. 2001,
and references therein). We note that values of the lithium opacity which
are higher than the ones we have used, are potentially possible. As an
extreme example, lithium abundance values as high as $X_{\rm LiI}\sim
10^{-8}$ were suggested by models of inhomogeneous Big--Bang
nucleosynthesis (Applegate \& Hogan 1985; Sale \& Mathews 1986; Mathews et
al. 1990),

The lithium signature on the CMB anisotropies provides the only direct
probe of structure in the universe at a redshift $z\sim 400$--500. This
redshift marks the beginning of the ``dark ages'' which end only after the
first generation of galaxies form at $z\sim 20$ (see review by Barkana \&
Loeb 2001).

\acknowledgements

We thank Daniel Eisenstein, David Hogg and Ur\v{o}s Seljak  
for useful discussions. This work was supported
in part by NASA grants NAG 5-7039, 5-7768, and by NSF grants AST-9900877,
AST-0071019 (for AL).

\newpage

\end{document}